\documentclass[12pt]{iopart}
\usepackage{graphicx}

\begin{document}

\title{Thermalization of quark-gluon matter by 2-to-2 and 3-to-3 elastic  
scatterings}

\author{Xiao-Ming Xu}

\address{Department of Physics, Shanghai University, Baoshan,
Shanghai 200444, China}
\ead{xmxu@xmxucao.sina.net}
\begin{abstract}
Thermalization of quark-gluon matter is studied with a transport equation that
includes contributions of 2-to-2 and 3-to-3 elastic scatterings. Thermalization
time is related to the squared amplitudes for the elastic scatterings that are
calculated in perturbative QCD.
\end{abstract}


\section{Introduction}
For a nonequilibrium partonic system, 
two questions must be answered. One is that
whether or not a thermal state can be established; another is that how fast
thermalization of the system goes.
The H-theorem ensures that the system evolves into a thermal state as a result 
of 2-body elastic scatterings \cite{reichl}. In a system where 3-body elastic 
scatterings are important, a transport equation needs a term that
reflects the 3-body elastic scatterings. 
\begin{eqnarray}
& & 
\frac {\partial f_1}{\partial t} 
+ {\rm\bf v}_1 \cdot {\nabla}_{\bf r} f_1 
         \nonumber    \\
& &
= -\frac {\rm g}{2E_1{\rm g}_{22}} \int \frac {d^3p_2}{(2\pi)^32E_2}
\frac {d^3p_3}{(2\pi)^32E_3} \frac {d^3p_4}{(2\pi)^32E_4}  
(2\pi)^4 \delta^4(p_1+p_2-p_3-p_4)
         \nonumber    \\
& &
~~~ \times \mid {\cal M}_{2 \to 2} \mid^2
[f_1f_2(1 \pm f_3)(1 \pm f_4)-f_3f_4(1 \pm f_1)(1 \pm f_2)]
         \nonumber    \\
& &
~~~ -\frac {{\rm g}^2}{2E_1{\rm g}_{33}} 
\int \frac {d^3p_2}{(2\pi)^32E_2}
\frac {d^3p_3}{(2\pi)^32E_3} \frac {d^3p_4}{(2\pi)^32E_4}  
\frac {d^3p_5}{(2\pi)^32E_5} \frac {d^3p_6}{(2\pi)^32E_6}  
         \nonumber    \\
& &
~~~ \times (2\pi)^4 \delta^4(p_1+p_2+p_3-p_4-p_5-p_6)
\mid {\cal M}_{3 \to 3} \mid^2 
         \nonumber    \\
& &
~~~ \times [f_1f_2f_3(1 \pm f_4)(1 \pm f_5)(1 \pm f_6)
           -f_4f_5f_6(1 \pm f_1)(1 \pm f_2)(1 \pm f_3)]
         \nonumber    \\
\end{eqnarray}
where $\rm g$ is the color-spin degeneracy factor, 
${\rm g}_{22}=n^\prime_{\rm out} !$ and 
${\rm g}_{33}=n_{\rm in} ! n_{\rm out} !$ where 
$n^\prime_{\rm out}$ ($n_{\rm out}$) is the number of identical final partons
of 2-to-2 (3-to-3) scatterings and $n_{\rm in}$ for the 3-to-3
scatterings is the number of 
identical initial partons except the parton in the distribution function 
$f_1$. $\mid {\cal M}_{2 \to 2} \mid^2$ and $\mid {\cal M}_{3 \to 3} \mid^2$
represent squared amplitudes of 2-to-2 and 3-to-3 parton
scatterings, respectively. 

Define 
\begin{equation}
H(t)=\int d\vec {r} d\vec {p}_1 f(\vec {p}_1, \vec {r}, t) 
\ln f(\vec {p}_1, \vec {r}, t).
\end{equation}
$H(t)$ is negative for $f<1$ and 
$H(t)$ always decreases with increasing time until it is independent of time
because $f$ obeys the transport equation.
When $H(t)$ does not rely on time, 
the distribution function takes a form of thermal distribution. Such reasoning 
proves that the system satisfying (1)
is driven toward a thermal state by both 2-body  
and 3-body elastic scatterings. For a system that is governed by
2-body and 3-body elastic scatterings, the first question is answered.

The answer to the second question depends on the squared amplitudes     
$\mid {\cal M}_{2 \to 2} \mid^2$ and $\mid {\cal M}_{3 \to 3} \mid^2$ which
relate to dynamical processes. The very simple, instructive but unrealistic 
case is that the squared amplitudes are contants. Even in the case 
the integrals on 
the right-hand side of the transport equation are generally not zero.  
An initial distribution function is anisotropic in momentum 
space for a nonequilibrium system. If the two squared amplitudes are zero,
the distribution function remains unchanged. Then thermalization is never
possible. If the two squared amplitudes take small values, the distribution 
function changes slowly with increasing time and thermalization is slow. 
Large values of the two squared amplitudes will cause large
change of the distribution function and the corresponding thermalization goes 
fast. Even though the squared amplitudes are constants, we still need to 
numerically solve the transport equation to get thermalization time which is
the difference of the time when the initial distribution is given and the 
time when the equation produces an isotropic momentum distribution.  
Therefore, how fast a system thermalizes depends on dynamical processes like
elastic scatterings and anisotropy of initial distribution.  

For a realistic case of interest, quark-gluon matter created in 
ultra-relativistic heavy-ion collisions, the squared amplitudes for 2-to-2 and
3-to-3 parton elastic scatterings have complicated dependences on parton 
momenta. Since the elastic scatterings take place among gluons, quarks and 
antiquarks with the same and/or different flavors, the study of how fast 
quark-gluon matter thermalizes becomes a very difficult task. However, from
the discussion in the last paragraph, we immediately realize that 
thermalization of quark-gluon matter goes faster and faster if more and more 
types of parton elastic scatterings are included in
$\mid {\cal M}_{2 \to 2} \mid^2$ and $\mid {\cal M}_{3 \to 3} \mid^2$.
A relevant change of thermalization time is shown by our recent works
\cite{xmxu1,xmxu2,xmxu3} that have taken into account the four types of
3-to-3 elastic scatterings: quark-quark-quark, 
quark-quark-antiquark and quark-antiquark-antiquark for quark matter and  
gluon-gluon-gluon for gluon matter.The 3-to-3 elastic scatterings, 
thermalization time and summary are given in the next two sections.

\section{3-to-3 Elastic scatterings}

The 3-to-3 quark scatterings at the lowest order are shown in figure 1
\cite{xmxu1}. The left diagram indicates
two-gluon-exchange induced scattering and the right involves triple-gluon 
coupling. Initial (final) quarks may have the same or different flavors. 
While two or three initial (final) quarks have the same flavor, exchanges
of quarks must be taken into account. 

Two typical processes of 3-to-3 quark-quark-antiquark scatterings at the tree 
level are shown in figure 2 \cite{xmxu2}. A lot of processes are related to 
annihilation and creation of quark and antiquark.
An exchange of the two initial (final) quarks needs to be considered if
the quarks have the same flavor. Quark-antiquark-antiquark elastic scatterings 
are similar to quark-quark-antiquark elastic scatterings.  

Because of triple-gluon couplings and four-gluon couplings, the  
3-to-3 gluon elastic scatterings are complicated. Four typical processes
at the lowest order are exhibited in figure 3 \cite{xmxu3}. Only the diagram 
${\rm B}_{\sim \sim}$ involves two four-gluon couplings. The other three 
diagrams include one four-gluon coupling and two triple-gluon couplings. The 
diagram ${\rm B}_{\rm U}$ has the four-gluon coupling related to two final 
gluons, ${\rm B}_{\sim +}$ to two final gluons and one initial gluon, and
${\rm B}_{\sim - 4(56)}$ to one exchanged final gluon and two initial gluons.  
     
Except for the diagram ${\rm B}_{\sim \sim}$ of which the squared amplitude is 
calculable by hand, we make fortran codes in Feynman gauge
to derive the squared amplitudes of
other diagrams that indicate the elastic scatterings of $qqq \to qqq$, 
$qq\bar {q} \to qq\bar {q}$, $q\bar {q}\bar {q} \to q\bar {q}\bar {q}$ and 
$ggg \to ggg$, respectively. While the squared
amplitudes for the 3-to-3 elastic scatterings are used in the transport
equation, the quark and gluon propagators are regularized with a screening 
mass \cite{bmw} to remove Coulomb exchange divergence.

\section{Numerical solutions and summary}

Anisotropic parton momentum distributions are formed in initial central Au-Au  
collisions. Such anisotropy can be eliminated by elastic scatterings among  
partons. 
For a central Au-Au collision at $\sqrt {s_{NN}}=200$ GeV, we obtain from
HIJING Monte Carlo simulation \cite{wwwgg,lmw}
the initial gluon distribution that corresponds 
to a gluon number density of 19.4 ${\rm fm}^{-3}$,
\begin{equation}
f(\vec {p})=\frac {17.1 (2\pi)^{1.5}}
{{\rm g}_{\rm G}\pi R_A^2 Y(\mid \vec {p} \mid/\cosh ({\rm y})+0.3)}
{\rm e}^{-\mid \vec {p} \mid/(0.9\cosh ({\rm y}))-(\mid \vec {p} \mid 
\tanh ({\rm y}))^2/8} 
\end{equation}
where $R_A=6.4$ fm, ${\rm g}_G=16$ and the rapidity $-5 \leq y \leq 5$. 
Any of the up-quark, down-quark, up-antiquark and down-antiquark initial
distribution functions is assumed to be
one sixth of the gluon initial distribution function.
Starting from the time $t_{\rm ini}$, when anisotropic quark-gluon matter  
is formed and ending at the time $t_{\rm iso}$, when local momentum isotropy is
established, the transport equation is solved. Local momentum isotropy,
in other words, a thermal state is
obtained if distribution function curves in different directions finally 
overlap through the thermalization time $t_{\rm iso}-t_{\rm ini}$.
For quark matter that is governed by the $qq \to qq$ and $qqq \to qqq$
elastic scatterings, thermalization 
time is about 1.8 fm/$c$; for quark matter interacted by antiquark
matter via  $q\bar {q} \to q\bar {q}$, $qq\bar {q} \to qq\bar {q}$ 
and $q\bar {q}\bar {q} \to q\bar {q}\bar {q}$ elastic scatterings, 
thermalization time is about 1.55 fm/$c$; for gluon matter the 
$gg \to gg$ and $ggg \to ggg$ elastic scatterings give a thermalization time 
of the order of 0.45 fm/$c$. While the $qq \to qq$ and $qqq \to qqq$ elastic
scatterings give a long thermalization time, the $q\bar {q} \to q\bar q$,
$qq\bar {q} \to qq\bar {q}$ and $q\bar {q}\bar {q} \to q\bar {q}\bar {q}$ 
elastic scatterings apparently shorten thermalization time of quark matter. 

In summary, the larger the squared amplitudes for the 2-to-2 and 3-to-3 
elastic scatterings are, the faster thermalization goes. 
$\mid {\cal M}_{3 \to 3} \mid^2$ for quark matter gets larger and larger
if we do a successive inclusion of quark-quark-quark, 
quark-quark-antiquark and quark-antiquark-antiquark elastic scatterings;
correspondingly, thermalization time of quark matter becomes shorter and 
shorter. Rapid thermalization of gluon matter is obtained from the
2-to-2 and 3-to-3 gluon elastic scatterings. The study of thermalization
leads to the importance of multi-parton elastic scatterings 
\cite{liuko} in addition to the 2-body parton scatterings 
\cite{sm,xg,wong,ndmg}. We note here that we have not studied the 
contributions of $qg \to qg$, $qqg \to qqg$,
$q\bar {q}g \to q\bar {q}g$ and $qgg \to qgg$
since our strategy is that we first give an independent study of gluon matter 
thermalization or of quark matter thermalization
and then study the influence of the interplay of quark
matter, antiquark matter and gluon matter. Because the derivation of the 
squared amplitude for a 3-to-3 elastic scattering is complicated and very 
time-consuming and numerically solving the transport equation needs months,  
we have to study the 3-to-3 elastic scatterings one by one. 
But the contributions of $qg \to qg$, $qqg \to qqg$,
$q\bar {q}g \to q\bar {q}g$ and $qgg \to qgg$ are expected to
shorten thermalization time of quark matter since they increase both
$\mid {\cal M}_{2 \to 2} \mid^2$ and $\mid {\cal M}_{3 \to 3} \mid^2$, and
the contributions will be considered in the coming years.

\ack
This work was supported by the National Natural Science Foundation of
China under Grant No. 10675079.

\newpage
\begin{figure}
  \centering
    \includegraphics[width=42mm,height=65mm,angle=0]{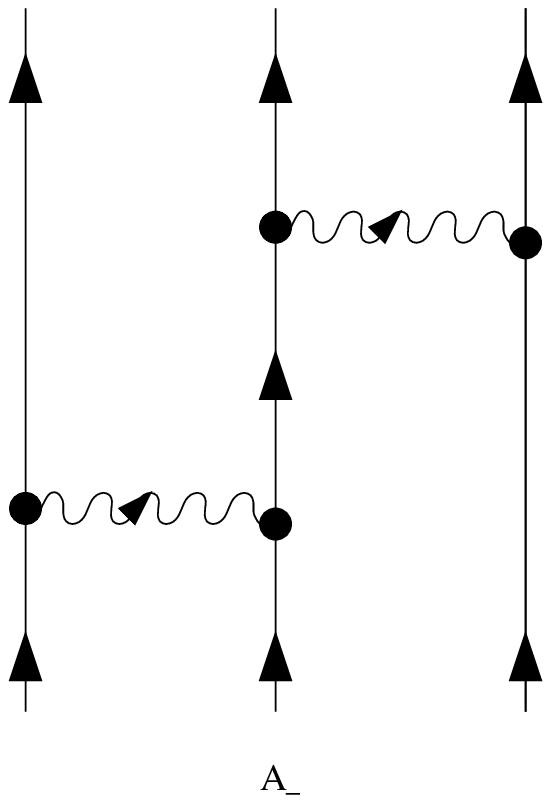}
      \hspace{1.2cm}
    \includegraphics[width=42mm,height=65mm,angle=0]{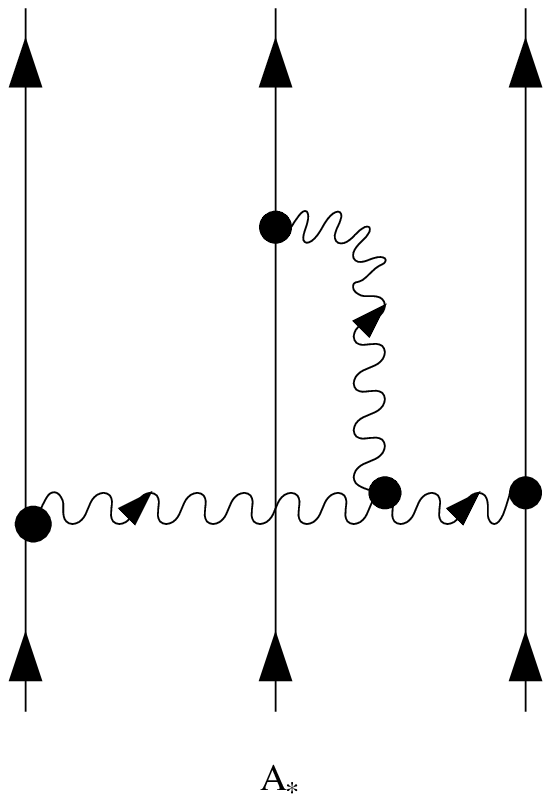}
      \hspace{1.2cm}
\caption{Quark-quark-quark elastic scatterings.}
\label{fig1}
\end{figure}

\newpage
\begin{figure}[t]
  \begin{center}
    \includegraphics[width=42mm,height=65mm,angle=0]{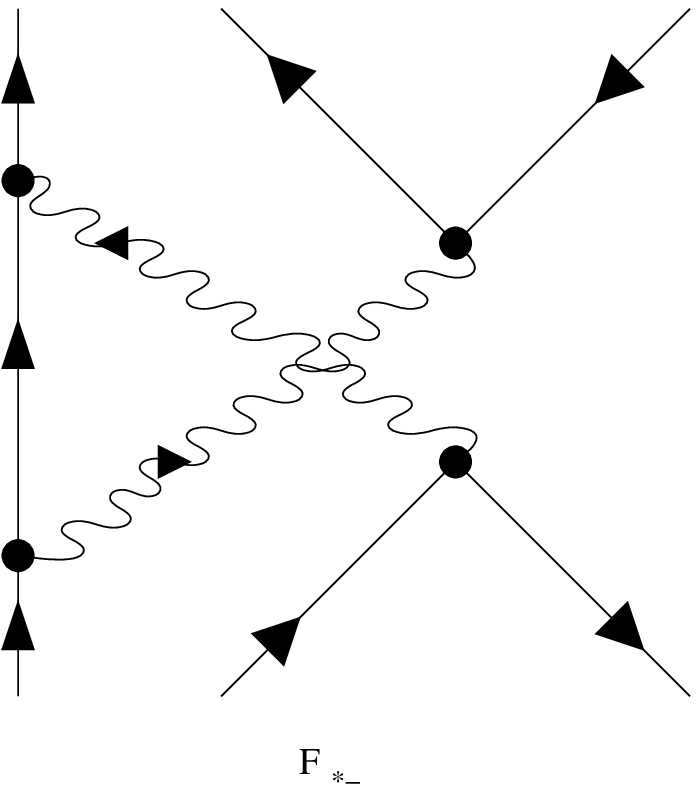}
      \hspace{1.2cm}
    \includegraphics[width=42mm,height=65mm,angle=0]{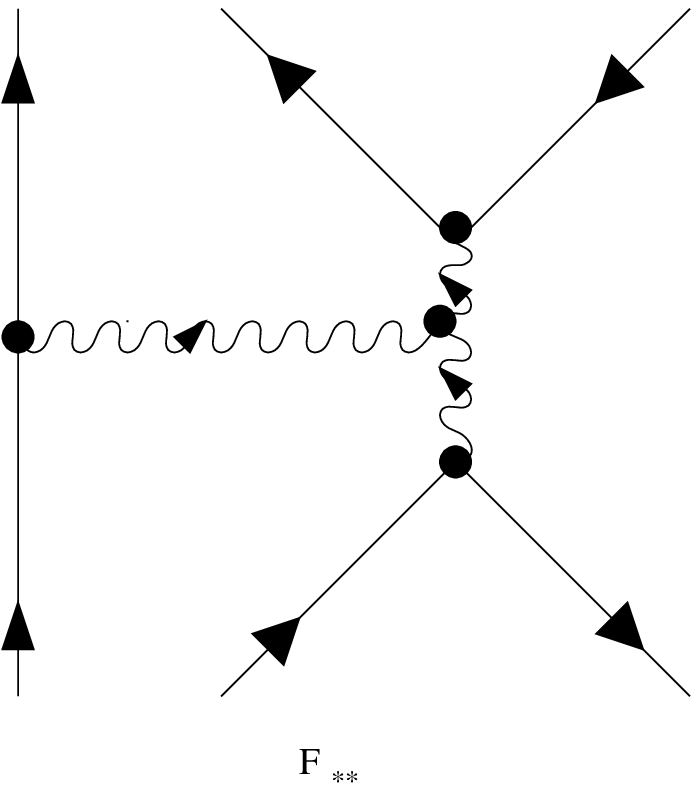}
      \hspace{1.2cm}
  \end{center}
\caption{Quark-quark-antiquark elastic scatterings.}
\label{fig2}
\end{figure}

\newpage
\begin{figure}[t]
  \begin{center}
    \leavevmode
    \includegraphics[width=42mm,height=65mm,angle=0]{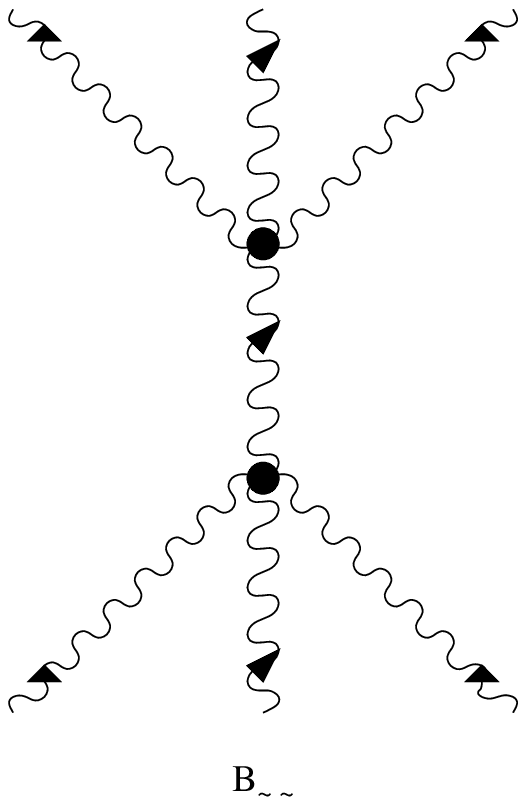}
      \hspace{1.2cm}
    \includegraphics[width=42mm,height=65mm,angle=0]{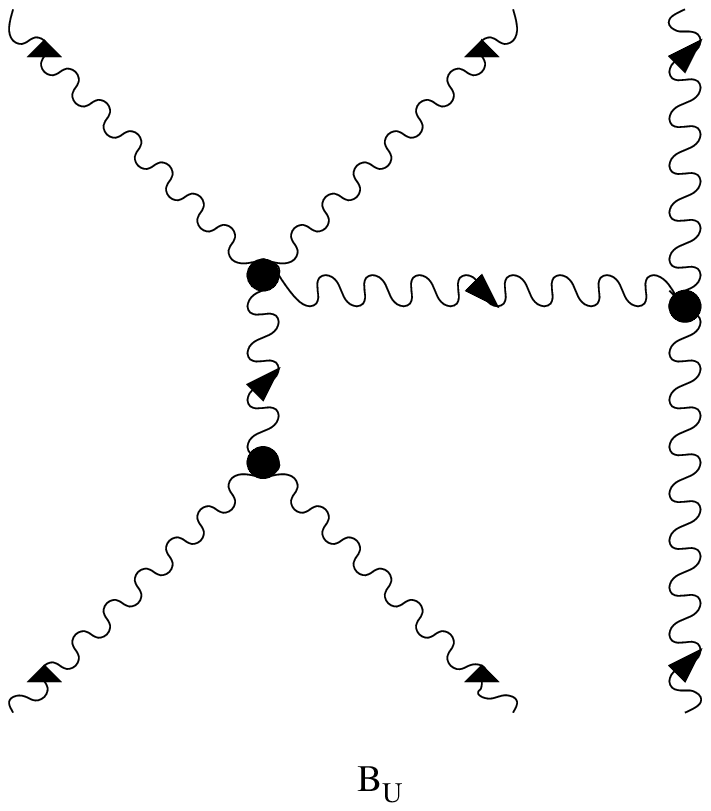}
      \vskip 26pt
    \includegraphics[width=42mm,height=65mm,angle=0]{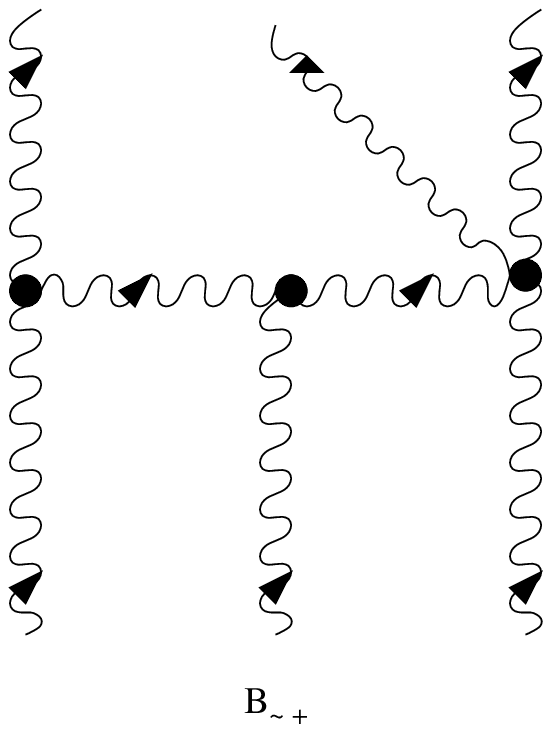}
      \hspace{1.2cm}
    \includegraphics[width=42mm,height=65mm,angle=0]{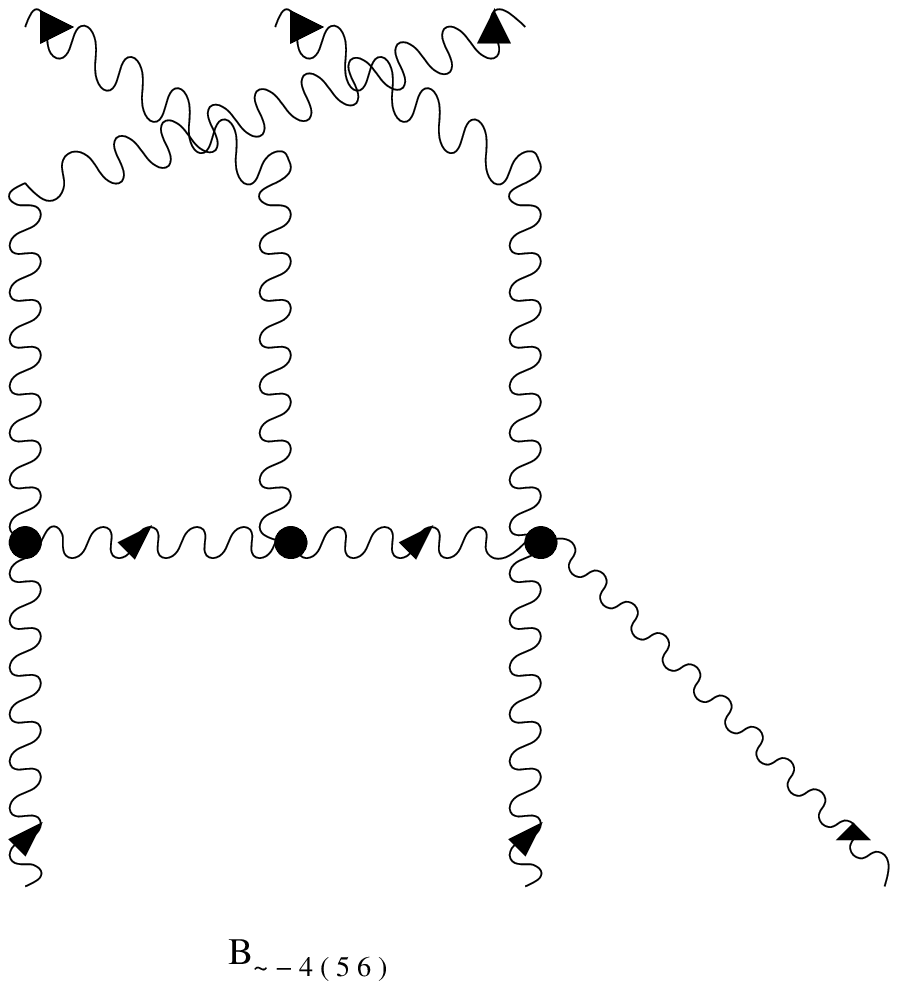}
      \vskip 26pt
  \end{center}
\caption{Gluon-gluon-gluon elastic scatterings.}
\label{fig3}
\end{figure}

\end{document}